\documentclass[aps,preprint]{revtex4}%
\usepackage{amsfonts}
\usepackage{amsmath}
\usepackage{amssymb}
\usepackage{graphicx}%
\begin{document}
\title[Coulomb Potential and Relativity]{Relativistic Mechanics and a Special Role for the Coulomb Potential}
\author{Timothy H. Boyer}
\affiliation{Department of Physics, City College of the City University of New York, New
York, New York 10031}

\begin{abstract}
It is shown that a nonrelativistic mechanical \ system involving a general
nonrelativistic potential $V(|\mathbf{r}_{1}-\mathbf{r}_{2}|)$ between point
particles at positions $\mathbf{r}_{1}$ and $\mathbf{r}_{2}$ can be extended
to a Lagrangian system which is invariant under Lorentz transformation through
order $v^{2}/c^{2}.$ \ However, this invariance requires the introduction of
velocity-dependent and acceleration-dependent forces between particles. \ The
textbook treatments of "relativistic mechanics" can be misleading; the
discussions usually deal with only one particle experiencing prescribed forces
and so make no mention of these additional velocity- and
acceleration-dependent forces. \ A simple example for a situation analogous to
a parallel-plate capacitor is analyzed for all the conservation laws of
Galilean invariance or Lorentz invariance. \ For this system, Galilean
invariance requires that the mechanical momentum is given by $\mathbf{p}%
_{mech}=m\mathbf{v}$ but places no restriction on the position-dependent
potential function. \ On the other hand, Lorentz invariance requires that the
mechanical momentum is given by $\mathbf{p}_{mech}=m\mathbf{v}(1-v^{2}%
/c^{2})^{-1/2},$ and in addition requires that the potential function is
exactly the Coulomb potential $V(|\mathbf{r}_{1}-\mathbf{r}_{2}%
|)=k/|\mathbf{r}_{1}-\mathbf{r}_{2}|.$ \ It is also noted that the
transmission of the interparticle-force signal at the speed of light again
suggests a special role for the Coulomb potential. \ A nonrelativistic
particle system interacting through the Coulomb potential becomes the Darwin
Lagrangian when extended to a system relativistic through order $v^{2}/c^{2},$
and then allows extension to classical electrodynamics as a fully
Lorentz-invariant theory of interacting particles.

\end{abstract}
\maketitle

\subsection{Introduction}

Contemporary physics regards special relativity as a metatheory to which
(locally) all theories describing nature should conform. \ Thus in
nonrelativistic classical mechanics, there is the unspoken implication that
the nonrelativistic interaction between point particles at positions
$\mathbf{r}_{1}$ and $\mathbf{r}_{2}$ under a general potential $V(|\mathbf{r}%
_{1}-\mathbf{r}_{2}|)$ is the small-velocity limit of some fully relativistic
theory of interacting point particles which might occur in nature. \ However,
the use of a general potential can be misleading for both students and
researchers. Here we demonstrate that an arbitrary nonrelativistic potential
function can indeed be extended to a Lagrangian which is Lorentz-invariant
through order $v^{2}/c^{2};$ however, the extension requires the introduction
of velocity-dependent and acceleration-dependent forces which go unmentioned
in the mechanics textbooks. Also, we present a simple example showing that the
$1/r$ potential between point particles is singled out as the only
nonrelativistic force law which will lead to appropriate Lorentz-invariant
behavior (\textit{without} the appearance of these additional forces) for
groups of particles arranged in a fashion analogous to parallel capacitor
plates. \ Finally, we note that the Coulomb potential is suggested when the
potential satisfies the wave equation for signal transmission at the speed of
light $c.$ All these results emphasize both the sometimes misleading nature of
current textbook treatments of "relativistic mechanics" and also the special
role played by the Coulomb potential.

\subsection{Textbook Discussion of the "Relativistic Lagrangian"}

Current textbooks of classical mechanics encourage the common misconception
among physicists that a \textit{relativistic} classical system can be obtained
from a \textit{nonrelativistic} mechanical system involving an arbitrary
nonrelativistic potential $V(|\mathbf{r}_{1}-\mathbf{r}_{2}|)$ between
particles simply by introducing the relativistic expressions for mechanical
linear momentum and mechanical energy for the particles. \ Thus, for example,
standard classical mechanics textbooks suggest\cite{Goldstein}\cite{Jose} that
the "relativistic Lagrangian" is obtained by using the relativistic Lagrangian
for a free particle and adding an arbitrary nonrelativistic potential. \ One
text\cite{Goldstein2} indeed has a section on "The relativistic
one-dimensional harmonic oscillator." \ 

The usual discussion of relativistic particle motion in classical mechanics
texts considers only a single particle $m~$and involves the replacement of the
Lagrangian $L(\mathbf{r,\dot{r})}$ for the nonrelativistic motion in a
time-independent potential $V(\mathbf{r})$
\begin{equation}
L(\mathbf{r},\mathbf{\dot{r})=}\frac{1}{2}m\mathbf{\dot{r}}^{2}-V(\mathbf{r)}%
\end{equation}
giving the nonrelativistic equation of motion%
\begin{equation}
\frac{d}{dt}(m\mathbf{\dot{r})=-\nabla}V\mathbf{(r)}%
\end{equation}
by the "relativistic Lagrangian"%
\begin{equation}
L(\mathbf{r},\mathbf{\dot{r})=}-mc^{2}(1-\mathbf{\dot{r}}^{2}/c^{2}%
)^{1/2}-V(\mathbf{r)}%
\end{equation}
with the equation of motion%
\begin{equation}
\frac{d}{dt}\left(  \frac{m\mathbf{\dot{r}}}{(1-\mathbf{\dot{r}}^{2}%
/c^{2})^{1/2}}\right)  \mathbf{=-\nabla}V\mathbf{(r)}%
\end{equation}
Thus in the equation of motion, the nonrelativistic particle momentum
$\mathbf{p}_{nonrel}=m\mathbf{\dot{r}}$ is replaced by the relativistic
particle momentum $\mathbf{p}_{rel}=m\mathbf{\dot{r}(}1-\mathbf{\dot{r}}%
^{2}/c^{2})^{-1/2}$, and the time rate of change of the momentum is given by
the same force $\mathbf{-\nabla}V\mathbf{(r)}$ in both relativistic and
nonrelativistic cases. \ Of course, these one-particle systems take this
simple form in only one inertial frame. \ In other inertial frames, there are
velocity-dependent and acceleration-dependent forces.

Some authors\cite{Brehme} go one step further and insist that the Lagrangian
itself should be written in manifestly covariant form despite the fact that
the forces on the particle may take a simple form in only one inertial frame.
\ Such one-particle systems (other than the free particle) exhibit neither
conservation of linear momentum nor constant motion of the center of energy,
both of which are expected in a Lorentz-invariant system. \ These one-particle
systems may provide mathematical exercises for students; however, with the
sole exception of electromagnetic forces, they are largely irrelevant to
physics as a description of nature, and indeed are misleading to students,
instructors, and researchers.\cite{Blanco} \ Insistence upon a covariant
appearance is a mere distraction, with no connection to nature.\ Indeed, as
was pointed out long ago by Kretchmann,\cite{Kretchman} \textit{any}
expression can be written in manifestly Lorentz-covariant notation, indeed in
general covariant notation. \ 

\subsection{Nonrelativistic Lagrangians for Particles and Lorentz-Invariant
Exension to Order $v^{2}/c^{2}$}

\ When describing nature, we regard the fundamental interactions as those
between point particles. \ Thus here we turn to the question as to what
potential functions $V(|\mathbf{r}_{1}-\mathbf{r}_{2}|)$ between two point
particles can be regarded as describing the nonrelativistic limit arising from
a fully Lorentz-invariant interaction between particles. \ We first try to
solve this problem by working backwards, trying to construct a relativistic
theory which produces a specific potential in the nonrelativistic limit.

The nonrelativistic mechanical behavior of two point particles interacting
through a potential $V(|\mathbf{r}_{1}-\mathbf{r}_{2}|)$ can be written in
terms of a Lagrangian%
\begin{equation}
L(\mathbf{r}_{1},\mathbf{r}_{2},\mathbf{\dot{r}}_{1},\mathbf{\dot{r}}%
_{2})=\frac{1}{2}m_{1}\mathbf{\dot{r}}_{1}^{2}+\frac{1}{2}m_{2}\mathbf{\dot
{r}}_{2}^{2}-V(|\mathbf{r}_{1}-\mathbf{r}_{2}|)
\end{equation}
The invariance of this Lagrangian under spacetime translations and spatial
rotations leads to the conservation laws for energy, linear momentum, and
angular momentum. \ The system is also invariant under Galilean
transformations where the generator of proper Galilean transformations is
given by the system total mass times the system center of mass.\cite{CV} \ In
order to extend this system to a Lorentz-invariant system, we must preserve
the invariance of the Lagrangian under spacetime translations \ and spatial
rotations while changing the system invariance under Galilean transformations
over to invariance under Lorentz transformations. \ The generator of Lorentz
transformations is the system total energy times the system center of
energy.\cite{CV} \ The first step in this transformation is the replacement of
the nonrelativistic expression for particle kinetic energy by the Lagrangian
for a relativistic free particle
\begin{equation}
\frac{1}{2}m\mathbf{\dot{r}}^{2}\rightarrow-mc^{2}(1-\mathbf{\dot{r}}%
^{2}/c^{2})^{1/2}%
\end{equation}
just as was done in moving from Eq. (1) to Eq. (3) above. \ With this
replacement, the nonrelativistic Lagrangian of Eq. (5) becomes now%
\begin{equation}
L(\mathbf{r}_{1},\mathbf{r}_{2},\mathbf{\dot{r}}_{1},\mathbf{\dot{r}}%
_{2})=-m_{1}c^{2}(1-\mathbf{\dot{r}}_{1}^{2}/c^{2})^{1/2}-m_{2}c^{2}%
(1-\mathbf{\dot{r}}_{2}^{2}/c^{2})^{1/2}-V(|\mathbf{r}_{1}-\mathbf{r}_{2}|)
\end{equation}
This Lagrangian, which is of the sort given in the mechanics
textbooks,\cite{ref12} will lead to relativistic expressions for particle
kinetic energy and particle linear momentum. \ It is invariant under spacetime
translations and spatial rotations. However, this system is not Lorentz
invariant. \ 

Since the energy and momentum of an isolated system form a Lorentz
four-vector, we expect the potential energy $V(|\mathbf{r}_{1}-\mathbf{r}%
_{2}|)$ to be related to momentum in a different inertial frame. Let us label
as $S$ the inertial frame in which the potential function $V(|\mathbf{r}%
_{1}-\mathbf{r}_{2}|)$ gives the nonrelativistic interaction of the particles.
\ Then when viewed from any inertial frame $S^{\prime}$ moving with constant
velocity with respect to the frame $S,$ we expect to find velocity-dependent
forces between the particles in addition to the position-dependent forces
found in the frame $S.$ \ If we require Lorentz invariance through order
$v^{2}/c^{2},$ then the velocity-dependent terms must appear in the Lagrangian
in any inertial frame. \ By working backwards from the requirement of Lorentz
invariance through order $v^{2}/c^{2},$ we find that the Lagrangian extended
from the nonrelativistic expression (5) can be written as%
\begin{align}
L(\mathbf{r}_{1},\mathbf{r}_{2},\mathbf{\dot{r}}_{1},\mathbf{\dot{r}}_{2})  &
=-m_{1}c^{2}(1-\mathbf{\dot{r}}_{1}^{2}/c^{2})^{1/2}-m_{2}c^{2}(1-\mathbf{\dot
{r}}_{2}^{2}/c^{2})^{1/2}-V(|\mathbf{r}_{1}-\mathbf{r}_{2}|)\nonumber\\
&  +\frac{1}{2}V(|\mathbf{r}_{1}-\mathbf{r}_{2}|)\frac{\mathbf{\dot{r}}%
_{1}\cdot\mathbf{\dot{r}}_{2}}{c^{2}}-\frac{1}{2}V^{\prime}(|\mathbf{r}%
_{1}-\mathbf{r}_{2}|)\frac{\mathbf{\dot{r}}_{1}\cdot(\mathbf{r}_{1}%
-\mathbf{r}_{2})\mathbf{\dot{r}}_{2}\cdot(\mathbf{r}_{1}-\mathbf{r}_{2}%
)}{c^{2}\left\vert \mathbf{r}_{1}-\mathbf{r}_{2}\right\vert }%
\end{align}
where $V^{\prime}(|\mathbf{r}_{1}-\mathbf{r}_{2}|)$ refers to the derivative
of the potential function with respect to its argument. \ 

We can check the Lorentz invariance of this Lagrangian through order
$v^{2}/c^{2}$ by showing that the system center of energy moves with constant
velocity through order $v^{2}/c^{2}.$ \ Indeed, we expect\cite{CofE}%
\begin{equation}
\frac{d}{dt}(U\overrightarrow{X})=c^{2}\mathbf{P}%
\end{equation}
where $U$ is the system energy, $\overrightarrow{X}$ is the system center of
energy, and $\mathbf{P}$ is the system linear momentum. \ The system energy
$U$ times the center of energy of the system $\overrightarrow{X}$ through
zero-order in $v/c$ is given by%
\begin{equation}
U\overrightarrow{X}=m_{1}(c^{2}+\frac{1}{2}\mathbf{\dot{r}}_{1}^{2}%
)\mathbf{r}_{1}+m_{2}(c^{2}+\frac{1}{2}\mathbf{\dot{r}}_{2}^{2})\mathbf{r}%
_{2}+V(|\mathbf{r}_{1}-\mathbf{r}_{2}|)\frac{(\mathbf{r}_{1}+\mathbf{r}_{2}%
)}{2}%
\end{equation}
corresponding to the restmass energy and kinetic energy of the two particles
located at their respective positions $\mathbf{r}_{1}$ and $\mathbf{r}_{2}$
plus the interaction potential energy located half way between the positions
of the two particles. \ Since the Lagrangian in Eq. (8) has no explicit time
dependence, the system energy $U$ is constant in time. \ Taking the time
derivative of Eq. (10), we find%
\begin{align}
\frac{d}{dt}(U\overrightarrow{X})  &  =U\frac{d\overrightarrow{X}}{dt}%
=m_{1}(c^{2}+\frac{1}{2}\mathbf{\dot{r}}_{1}^{2})\mathbf{\dot{r}}_{1}%
+m_{2}(c^{2}+\frac{1}{2}\mathbf{\dot{r}}_{2}^{2})\mathbf{\dot{r}}_{2}%
+(m_{1}\mathbf{\ddot{r}}_{1}\mathbf{\cdot\dot{r}}_{1})\mathbf{r}_{1}%
+(m_{2}\mathbf{\ddot{r}}_{2}\mathbf{\cdot\dot{r}}_{2})\mathbf{r}%
_{2}\nonumber\\
&  +\frac{1}{2}V(|\mathbf{r}_{1}-\mathbf{r}_{2}|)(\mathbf{\dot{r}}%
_{1}\mathbf{+\dot{r}}_{2}\mathbf{)+}\frac{1}{2}V^{\prime}(|\mathbf{r}%
_{1}-\mathbf{r}_{2}|)\frac{(\mathbf{\dot{r}}_{1}-\mathbf{\dot{r}}_{2}%
)}{|\mathbf{r}_{1}-\mathbf{r}_{2}|}\cdot(\mathbf{r}_{1}-\mathbf{r}%
_{2})(\mathbf{r}_{1}+\mathbf{r}_{2})
\end{align}
It is sufficient to use the nonrelativistic equations of motion,%
\begin{equation}
m_{1}\mathbf{\ddot{r}}_{1}=-V^{\prime}(|\mathbf{r}_{1}-\mathbf{r}_{2}%
|)\frac{(\mathbf{r}_{1}-\mathbf{r}_{2})}{\left\vert \mathbf{r}_{1}%
-\mathbf{r}_{2}\right\vert }%
\end{equation}%
\begin{equation}
m_{2}\mathbf{\ddot{r}}_{2}=V^{\prime}(|\mathbf{r}_{1}-\mathbf{r}_{2}%
|)\frac{(\mathbf{r}_{1}-\mathbf{r}_{2})}{\left\vert \mathbf{r}_{1}%
-\mathbf{r}_{2}\right\vert }%
\end{equation}
to transform Eq. (11) into the form%
\begin{align}
\frac{d}{dt}(U\overrightarrow{X})  &  =U\frac{d\overrightarrow{X}}{dt}%
=m_{1}(c^{2}+\frac{1}{2}\mathbf{\dot{r}}_{1}^{2})\mathbf{\dot{r}}_{1}%
+m_{2}(c^{2}+\frac{1}{2}\mathbf{\dot{r}}_{2}^{2})\mathbf{\dot{r}}%
_{2}\nonumber\\
&  -\left(  V^{\prime}(|\mathbf{r}_{1}-\mathbf{r}_{2}|)\frac{(\mathbf{r}%
_{1}-\mathbf{r}_{2})}{\left\vert \mathbf{r}_{1}-\mathbf{r}_{2}\right\vert
}\mathbf{\cdot\dot{r}}_{1}\right)  \mathbf{r}_{1}+\left(  V^{\prime
}(|\mathbf{r}_{1}-\mathbf{r}_{2}|)\frac{(\mathbf{r}_{1}-\mathbf{r}_{2}%
)}{\left\vert \mathbf{r}_{1}-\mathbf{r}_{2}\right\vert }\mathbf{\cdot\dot{r}%
}_{2}\right)  \mathbf{r}_{2}\nonumber\\
&  +\frac{1}{2}V(|\mathbf{r}_{1}-\mathbf{r}_{2}|)(\mathbf{\dot{r}}%
_{1}\mathbf{+\dot{r}}_{2}\mathbf{)+}\frac{1}{2}V^{\prime}(|\mathbf{r}%
_{1}-\mathbf{r}_{2}|)\frac{(\mathbf{\dot{r}}_{1}-\mathbf{\dot{r}}_{2}%
)}{|\mathbf{r}_{1}-\mathbf{r}_{2}|}\cdot(\mathbf{r}_{1}-\mathbf{r}%
_{2})(\mathbf{r}_{1}+\mathbf{r}_{2})
\end{align}
The momenta can be obtained from the Lagrangian in Eq. (8) as%
\begin{align}
\mathbf{p}_{1}  &  =\frac{\partial L}{\partial\mathbf{\dot{r}}_{1}}%
=m_{1}\mathbf{\dot{r}}_{1}\left(  1-\frac{\mathbf{\dot{r}}_{1}^{2}}{c^{2}%
}\right)  ^{-1/2}+\frac{1}{2}V(|\mathbf{r}_{1}-\mathbf{r}_{2}|)\frac
{\mathbf{\dot{r}}_{2}}{c^{2}}\nonumber\\
&  -\frac{1}{2}V^{\prime}(|\mathbf{r}_{1}-\mathbf{r}_{2}|)\frac{(\mathbf{r}%
_{1}-\mathbf{r}_{2})}{\left\vert \mathbf{r}_{1}-\mathbf{r}_{2}\right\vert
}\frac{\mathbf{\dot{r}}_{2}}{c^{2}}\cdot(\mathbf{r}_{1}-\mathbf{r}_{2})
\end{align}%
\begin{align}
\mathbf{p}_{2}  &  =\frac{\partial L}{\partial\mathbf{\dot{r}}_{2}}%
=m_{2}\mathbf{\dot{r}}_{2}\left(  1-\frac{\mathbf{\dot{r}}_{2}^{2}}{c^{2}%
}\right)  ^{-1/2}+\frac{1}{2}V(|\mathbf{r}_{1}-\mathbf{r}_{2}|)\frac
{\mathbf{\dot{r}}_{1}}{c^{2}}\nonumber\\
&  -\frac{1}{2}V^{\prime}(|\mathbf{r}_{1}-\mathbf{r}_{2}|)\frac{(\mathbf{r}%
_{1}-\mathbf{r}_{2})}{\left\vert \mathbf{r}_{1}-\mathbf{r}_{2}\right\vert
}\frac{\mathbf{\dot{r}}_{1}}{c^{2}}\cdot(\mathbf{r}_{1}-\mathbf{r}_{2})
\end{align}
giving total linear momentum%
\begin{align}
\mathbf{P}  &  \mathbf{=}m_{1}\mathbf{\dot{r}}_{1}\left(  1-\frac
{\mathbf{\dot{r}}_{1}^{2}}{c^{2}}\right)  ^{-1/2}+m_{2}\mathbf{\dot{r}}%
_{2}\left(  1-\frac{\mathbf{\dot{r}}_{2}^{2}}{c^{2}}\right)  ^{-1/2}%
\nonumber\\
&  +\frac{1}{2}V(|\mathbf{r}_{1}-\mathbf{r}_{2}|)\frac{\mathbf{\dot{r}}_{1}%
}{c^{2}}+\frac{1}{2}V(|\mathbf{r}_{1}-\mathbf{r}_{2}|)\frac{\mathbf{\dot{r}%
}_{2}}{c^{2}}\nonumber\\
&  -\frac{1}{2}V^{\prime}(|\mathbf{r}_{1}-\mathbf{r}_{2}|)\frac{(\mathbf{r}%
_{1}-\mathbf{r}_{2})}{\left\vert \mathbf{r}_{1}-\mathbf{r}_{2}\right\vert
}\frac{\mathbf{\dot{r}}_{1}}{c^{2}}\cdot(\mathbf{r}_{1}-\mathbf{r}_{2}%
)-\frac{1}{2}V^{\prime}(|\mathbf{r}_{1}-\mathbf{r}_{2}|)\frac{(\mathbf{r}%
_{1}-\mathbf{r}_{2})}{\left\vert \mathbf{r}_{1}-\mathbf{r}_{2}\right\vert
}\frac{\mathbf{\dot{r}}_{2}}{c^{2}}\cdot(\mathbf{r}_{1}-\mathbf{r}_{2})
\end{align}
Comparing Eqs. (14) and (17) after reorganizing a few terms, we find that
indeed Eq. (9) holds. \ The system of Eq. (8) is indeed Lorentz invariant
through order $v^{2}/c^{2}.$

\subsection{Velocity-Dependent and Acceleration-Dependent Forces in
Lorentiz-Invariant Systems}

The Lagrange equations of motion follow from the Lagrangian in Eq. (8); for
the particle at $\mathbf{r}_{1}$, the equation takes the form
\begin{align}
0 &  =\frac{d}{dt}\left(  \frac{m_{1}\mathbf{\dot{r}}_{1}}{(1-\mathbf{\dot{r}%
}_{1}^{2}/c^{2})^{1/2}}+\frac{1}{2}V(|\mathbf{r}_{1}-\mathbf{r}_{2}%
|)\frac{\mathbf{\dot{r}}_{2}}{c^{2}}-\frac{1}{2}V^{\prime}(|\mathbf{r}%
_{1}-\mathbf{r}_{2}|)\frac{(\mathbf{r}_{1}-\mathbf{r}_{2})}{\left\vert
\mathbf{r}_{1}-\mathbf{r}_{2}\right\vert }\frac{\mathbf{\dot{r}}_{2}}{c^{2}%
}\cdot(\mathbf{r}_{1}-\mathbf{r}_{2})\right)  \nonumber\\
&  -\frac{\mathbf{r}_{1}-\mathbf{r}_{2}}{\left\vert \mathbf{r}_{1}%
-\mathbf{r}_{2}\right\vert }V^{\prime}(|\mathbf{r}_{1}-\mathbf{r}_{2}|)\left(
-1+\frac{\mathbf{\dot{r}}_{1}\cdot\mathbf{\dot{r}}_{2}}{2c^{2}})+\frac
{\mathbf{\dot{r}}_{1}\cdot(\mathbf{r}_{1}-\mathbf{r}_{2})\mathbf{\dot{r}}%
_{2}\cdot(\mathbf{r}_{1}-\mathbf{r}_{2})}{2c^{2}\left\vert \mathbf{r}%
_{1}-\mathbf{r}_{2}\right\vert ^{2}}\right)  \text{ \ }\nonumber\\
&  +\frac{\mathbf{r}_{1}-\mathbf{r}_{2}}{\left\vert \mathbf{r}_{1}%
-\mathbf{r}_{2}\right\vert }V^{^{\prime\prime}}(|\mathbf{r}_{1}-\mathbf{r}%
_{2}|)\frac{\mathbf{\dot{r}}_{1}\cdot(\mathbf{r}_{1}-\mathbf{r}_{2}%
)\mathbf{\dot{r}}_{2}\cdot(\mathbf{r}_{1}-\mathbf{r}_{2})}{2c^{2}\left\vert
\mathbf{r}_{1}-\mathbf{r}_{2}\right\vert ^{2}}\nonumber\\
&  +V^{\prime}(|\mathbf{r}_{1}-\mathbf{r}_{2}|)\left(  \frac{\mathbf{\dot{r}%
}_{1}\cdot(\mathbf{r}_{1}-\mathbf{r}_{2})\mathbf{\dot{r}}_{2}+\mathbf{\dot{r}%
}_{2}\cdot(\mathbf{r}_{1}-\mathbf{r}_{2})\mathbf{\dot{r}}_{1}}{2c^{2}%
\left\vert \mathbf{r}_{1}-\mathbf{r}_{2}\right\vert }\right)
\end{align}
The equations of motion can be rewritten as forces acting on the particles to
change the mechanical momentum. \ For the particle at $\mathbf{r}_{1}$, this
becomes%
\begin{align}
\frac{d}{dt}\left(  \frac{m\mathbf{\dot{r}}_{1}}{(1-\mathbf{\dot{r}}_{1}%
^{2}/c^{2})^{1/2}}\right)    & =-V^{\prime}(|\mathbf{r}_{1}-\mathbf{r}%
_{2}|)\frac{\mathbf{r}_{1}-\mathbf{r}_{2}}{\left\vert \mathbf{r}%
_{1}-\mathbf{r}_{2}\right\vert }\left[  1+\frac{1}{2}\left(  \frac
{\mathbf{\dot{r}}_{2}}{c}\right)  ^{2}\right]  \nonumber\\
& -\frac{\mathbf{r}_{1}-\mathbf{r}_{2}}{2c^{2}\left\vert \mathbf{r}%
_{1}-\mathbf{r}_{2}\right\vert }\left(  \frac{V^{^{\prime\prime}}%
(|\mathbf{r}_{1}-\mathbf{r}_{2}|)}{\left\vert \mathbf{r}_{1}-\mathbf{r}%
_{2}\right\vert }-\frac{V^{\prime}(|\mathbf{r}_{1}-\mathbf{r}_{2}%
|)}{\left\vert \mathbf{r}_{1}-\mathbf{r}_{2}\right\vert ^{2}}\right)  \left[
\mathbf{\dot{r}}_{2}\cdot(\mathbf{r}_{1}-\mathbf{r}_{2})\right]
^{2}\nonumber\\
& -\frac{1}{2c^{2}}\left(  V(|\mathbf{r}_{1}-\mathbf{r}_{2}|)\mathbf{\ddot{r}%
}_{2}-V^{\prime}(|\mathbf{r}_{1}-\mathbf{r}_{2}|)\frac{[\mathbf{\ddot{r}}%
_{2}\cdot(\mathbf{r}_{1}-\mathbf{r}_{2})](\mathbf{r}_{1}-\mathbf{r}_{2}%
)}{|\mathbf{r}_{1}-\mathbf{r}_{2}|^{2}}\right)  \nonumber\\
& -\frac{\mathbf{\dot{r}}_{1}}{c}\times\left(  \frac{\mathbf{\dot{r}}_{2}}%
{c}\times\frac{\mathbf{r}_{1}-\mathbf{r}_{2}}{\left\vert \mathbf{r}%
_{1}-\mathbf{r}_{2}\right\vert }V^{\prime}(|\mathbf{r}_{1}-\mathbf{r}%
_{2}|)\right)
\end{align}
We notice that the force on the first particle involves not only the force
arising from the original nonrelativistic potential function, but also forces
depending upon the velocities of both particles and upon the acceleration of
the other particle. These forces were not part of the original nonrelativistic
theory. \ Such forces are absent from the accounts in the mechanics textbooks
and from the articles which treat "relativistic" motion for a single article.
The single particle appearing in the Lagrangian of these treatments produces
velocity-dependent and acceleration-dependent forces back on the prescribed
sources whose momentum and energy are never discussed.

The most famous Lagrangian which is Lorentz invariant through $v^{2}/c^{2}$ is
that obtained from the Coulomb potential $V(|\mathbf{r}_{1}-\mathbf{r}%
_{2}|)=q_{1}q_{2}/|\mathbf{r}_{1}-\mathbf{r}_{2}|.$ \ In this case the
Lagrangian of Eq. (8) becomes
\begin{align}
L(\mathbf{r}_{1},\mathbf{r}_{2},\mathbf{\dot{r}}_{1},\mathbf{\dot{r}}_{2}) &
=-m_{1}c^{2}(1-\mathbf{\dot{r}}_{1}^{2}/c^{2})^{1/2}-m_{2}c^{2}(1-\mathbf{\dot
{r}}_{2}^{2}/c^{2})^{1/2}-\frac{q_{1}q_{2}}{|\mathbf{r}_{1}-\mathbf{r}_{2}%
|}\nonumber\\
&  +\frac{1}{2}\frac{q_{1}q_{2}}{|\mathbf{r}_{1}-\mathbf{r}_{2}|}%
\frac{\mathbf{\dot{r}}_{1}\cdot\mathbf{\dot{r}}_{2}}{c^{2}}+\frac{1}{2}%
\frac{q_{1}q_{2}}{|\mathbf{r}_{1}-\mathbf{r}_{2}|}\frac{\mathbf{\dot{r}}%
_{1}\cdot(\mathbf{r}_{1}-\mathbf{r}_{2})\mathbf{\dot{r}}_{2}\cdot
(\mathbf{r}_{1}-\mathbf{r}_{2})}{c^{2}\left\vert \mathbf{r}_{1}-\mathbf{r}%
_{2}\right\vert ^{2}}%
\end{align}
If in Eq. (20) we expand the free-particle expressions $-mc^{2}(1-\mathbf{\dot
{r}}^{2}/c^{2})^{1/2}$ through second order in $v/c,$ then this becomes the
Darwin Lagrangian which sometime appears in electromagnetism textbooks as an
approximation to the interaction of charged particles.\cite{Jack} \ The
approximation is an accurate description of the classical electromagnetic
interaction between charged particles through second order in $v/c$ for small
separations between the particles. \ The Lagrangian equation of motion
following from Eq.(20) becomes (for the particle at position $\mathbf{r}_{1}$)%
\begin{align}
\frac{d}{dt}\left(  \frac{m_{1}\mathbf{\dot{r}}_{1}}{(1-\mathbf{\dot{r}}%
_{1}^{2}/c^{2})^{1/2}}\right)   &  =q_{1}[q_{2}\frac{(\mathbf{r}%
_{1}-\mathbf{r}_{2})}{|\mathbf{r}_{1}-\mathbf{r}_{2}|^{3}}\left\{  1+\frac
{1}{2}\left(  \frac{\mathbf{\dot{r}}_{2}}{c}\right)  ^{2}-\frac{3}{2}\left(
\frac{(\mathbf{r}_{1}-\mathbf{r}_{2})\cdot\mathbf{\dot{r}}_{2}}{|\mathbf{r}%
_{1}-\mathbf{r}_{2}|}\right)  ^{2}\right\}  \nonumber\\
&  -\frac{q_{2}}{2c}\left(  \mathbf{\ddot{r}}_{2}\mathbf{+}\frac
{[\mathbf{\ddot{r}}_{2}\cdot(\mathbf{r}_{1}-\mathbf{r}_{2})](\mathbf{r}%
_{1}-\mathbf{r}_{2})}{|\mathbf{r}_{1}-\mathbf{r}_{2}|^{2}}\right)
]\nonumber\\
&  +q_{1}\frac{\mathbf{\dot{r}}_{1}}{c}\times\left[  q_{2}\frac{\mathbf{\dot
{r}}_{2}}{c}\times\frac{(\mathbf{r}_{1}-\mathbf{r}_{2})}{|\mathbf{r}%
_{1}-\mathbf{r}_{2}|^{3}}\right]
\end{align}
where we have rewritten the Lagrangian equation in the form $d\mathbf{p}%
_{1}/dt=q_{1}\mathbf{E+}q_{1}(\mathbf{\dot{r}}_{1}/c)\times\mathbf{B}$ with
$\mathbf{p}_{1}$ the mechanical particle momentum.\cite{PA} \ The velocity-
and acceleration-dependent forces in Eq.(21) correspond to fields arising from
electromagnetic induction. \ In the textbooks, electromagnetic induction is
always treated without reference to any charged particles which may be
producing the induction fields, a very different point of view from that which
follows from the Darwin Lagrangian.

\subsection{Special Role for the Coulomb Potential for a Parallel-Disk System}

Since the potential energy $V(|\mathbf{r}_{1}-\mathbf{r}_{2}|)$ and associated
relativistic momentum depend upon pairs of particles whereas the mechanical
energy and momentum depend upon individual particles, the quantities
associated with the potential can be made arbitrarily large compared to the
mechanical quantities by considering a group of particles held together by
forces of constraint. \ This means that for a Lorentz-invariant multiparticle
system, the velocity-dependent and acceleration-dependent forces might
dominate any consideration of particle mechanics. \ And indeed this does occur
in connection with the self inductance and mutual inductance of
electromagnetic inductors where the mass of the charge carriers plays so small
a role that it is never mentioned. \ However, there is at least one case where
the additional velocity-dependent and acceleration-dependent forces do
\textit{not} dominate the multi-particle system. \ In an earlier
article\cite{Illustrations} providing illustrations of the center-of-energy
motion in relativistic systems, the example of a parallel-plate capacitor was
used. \ Here we point out that a Coulomb potential and only a Coulomb
potential allows relativistic behavior for a parallel-plate system without any
consideration of the velocity-dependent or acceleration-dependent forces
generally required for relativistic behavior.

The forces of constraint holding together a group of particles can introduce
energy and momentum into the system unless they are perpendicular to the
direction of motion of the group of particles. \ Because of this
consideration, we will consider two groups of interacting particles which are
arranged in a disk fashion analogous to those of a parallel-plate capacitor,
and we will consider the motion of the disks along a single axis perpendicular
to the plates. \ In this case, the forces of constraint holding each plate
together are perpendicular to the direction of motion and so introduce neither
energy nor momentum into the system.

In order to calculate the forces between the plates, we sum the forces between
particles assuming superposition holds. \ For definiteness, we assume a
potential of the form $1/r^{n}$, so that the force between two particles A and
B in the nonrelativistic limit takes the form
\begin{equation}
\mathbf{F}_{onA}=-\nabla_{A}V(|\mathbf{r}_{A}-\mathbf{r}_{B}|)=\frac
{-nk(\mathbf{r}_{A}-\mathbf{r}_{B})}{|\mathbf{r}_{A}-\mathbf{r}_{B}|^{n+2}}%
\end{equation}
where
\begin{equation}
V(|\mathbf{r}_{A}-\mathbf{r}_{B}|)=\frac{-k}{|\mathbf{r}_{A}-\mathbf{r}%
_{B}|^{n}}%
\end{equation}
and we assume $n>0$ so that the potential decreases as the separation between
the particles increases. \ Next we consider a uniform disk of particles of
type $B$ in the $yz$-plane with $\sigma$ particles per unit area. \ We obtain
the force on particle $A$ at a small distance $L\,\ $above the center of the
disk of large radius $R,$ $R>>L$, by summing the contributions of the
particles in the disk. \ Taking account of the cylindrical symmetry, the force
on $A$ is given by%
\begin{align}
\mathbf{F}_{onA}  &  =%
{\displaystyle\int\limits_{0}^{R}}
\left(  2\pi rdr\right)  \sigma_{B}nk\frac{-L}{(L^{2}+r^{2})^{n/2+1}%
}\nonumber\\
&  =2\pi\sigma_{B}kL\left(  \frac{1}{(L^{2}+r^{2})^{n/2}}\right)  _{r=0}%
^{r=R}\nonumber\\
&  =-2\pi\sigma_{B}k\left(  \frac{1}{L^{n-1}}\right)
\end{align}
where we have used the assumptions $n>0$ and $R>>L.$ \ From this result, we
can obtain the attractive forces between a pair of parallel disks of radius
$R,$ one made up of particles of type $A$ and the other of type $B$, separated
by a distance $L$,
\begin{equation}
\mathbf{F}=-2\pi\sigma_{B}kL^{1-n}(\sigma_{A}\pi R^{2})=-2\pi^{2}\sigma
_{A}\sigma_{B}kR^{2}L^{1-n}%
\end{equation}
The potential energy function associated with this force is
\begin{equation}
V(L)=2\pi^{2}\sigma_{A}\sigma_{B}kR^{2}\frac{L^{2-n}}{2-n}%
\end{equation}

We now go over to the mechanical motion. \ We imagine that the two disks are
allowed to accelerate toward each other due to the force between them. \ We
wish to consider the conservation laws for the system of these two disks. \ We
imagine the two disks as being oriented parallel to the $yz$-plane with the
$x$-axis running through the centers of the disks. \ The disk of particles of
type $A$ has mass $m$ and is located at $x$ and while the other disk has mass
$M$ and is located at coordinate $X$ with $x<X.$ If we write the constants
appearing in Eqs. (25) and (26) as%
\begin{equation}
C=2\pi^{2}\sigma_{A}\sigma_{B}kR^{2}%
\end{equation}
and assume that this force is the only force acting on the disks, then
Newton's equations of motion for the disks give the momentum changes along the
$x$-axis as%
\begin{equation}
\frac{dp_{m}}{dt}=C(X-x)^{1-n}=-\frac{dp_{M}}{dt}%
\end{equation}

If we assume that the system momentum is entirely mechanical, then the total
momentum $\mathbf{P}$ is given by%

\begin{equation}
\mathbf{P=}\widehat{i}p_{m}+\widehat{i}p_{M}%
\end{equation}
By symmetry, the angular momentum taken about the origin vanishes%
\begin{equation}
\mathbf{L=}0.
\end{equation}
The total energy $U$ of the system includes both the mechanical energies
$U_{m}$ and $U_{M}$ of the disks and the potential energy $V$ between the
disks as given in Eq. (26)%
\begin{equation}
U=U_{m}+U_{M}+\frac{C}{2-n}(X-x)^{2-n}%
\end{equation}
The center of rest-mass $\overrightarrow{\mathcal{X}}_{mass}$ of the system is
given by
\begin{equation}
(m+M)\overrightarrow{\mathcal{X}}_{mass}=m\widehat{i}x+M\widehat{i}X
\end{equation}
while the center of energy $\overrightarrow{\mathcal{X}}_{energy}$ of the
system is given by%
\begin{equation}
U\overrightarrow{\mathcal{X}}_{energy}=U_{m}\widehat{i}x+U_{M}\widehat
{i}X+\frac{C}{2-n}(X-x)^{2-n}\frac{(\widehat{i}x+\widehat{i}X)}{2}%
\end{equation}
where the center-of-energy location for the potential energy has been taken as
half-way between the disks.

We now wish to consider the conservation laws for this system. \ The
conservation of linear momentum associated with space-translation invariance
in the $x$-direction follows as%
\begin{equation}
\frac{d\mathbf{P}}{dt}=\widehat{i}\frac{dp_{m}}{dt}+\widehat{i}\frac{dp_{M}%
}{dt}=0
\end{equation}
from Newton's equations of motion in Eq. (28). \ The conservation of energy
associated with time-translation invariance follows as%
\begin{align}
\frac{dU}{dt}  &  =\frac{dU_{m}}{dt}+\frac{dU_{M}}{dt}+\frac{dV}%
{dt}\nonumber\\
&  =\left(  \frac{dp_{m}}{dt}-C(X-x)^{1-n}\right)  \frac{dx}{dt}+\left(
\frac{dp_{M}}{dt}+C(X-x)^{1-n}\right)  \frac{dX}{dt}=0
\end{align}
where we have used $dV/dt=C(X-x)^{1-n}(dx/dt+dX/dt)$ together with the
equations of motion appearing in Eq. (28) and the basic relation%
\begin{equation}
\frac{dU_{mech}}{dt}=\frac{d\mathbf{p}_{mech}}{dt}\cdot\frac{d\mathbf{r}}{dt}%
\end{equation}
which holds for each disk.

When considering the conservation of linear momentum, angular momentum, and
energy, we have not had to specify whether our system was invariant under
Galilean transformations or under Lorentz transformations. \ However, we now
wish to apply the last conservation law associated with change from one
inertial frame to another. \ The generator of Galilean
transformations\cite{CV} is the total rest-mass times the center of rest mass
as in Eq. (32). \ The conservation law associated with this generator is
related to the relativistic conservation law in Eq. (9) when we divide Eq. (9)
by $c^{2}$ and take the limit $c^{2}\rightarrow\infty;$ this leaves only the
restmass contributions to the energy and the linear momentum%
\begin{equation}
\frac{d}{dt}\left(  (\Sigma_{i}m_{i})\overrightarrow{\mathcal{X}}%
_{mass}\right)  =\Sigma_{i}\left(  m_{i}\frac{d\mathbf{r}_{i}}{dt}\right)
=\mathbf{P}%
\end{equation}
For our example involving Eq. (32), this gives the conservation law%
\begin{align}
\frac{d}{dt}((m+M)\overrightarrow{\mathcal{X}}_{mass}) &  =\frac{d}%
{dt}(m\widehat{i}x+M\widehat{i}X)\nonumber\\
&  =\widehat{i}m\frac{dx}{dt}+\widehat{i}M\frac{dX}{dt}=\mathbf{P}%
\end{align}
If we compare this Eq. (38) with Eq. (29) for the total momentum, we see that
we must identify%
\begin{equation}
\mathbf{p}_{m}=\widehat{i}m\frac{dx}{dt}\text{ \ \ \ \ }\mathbf{p}%
_{M}=\widehat{i}M\frac{dX}{dt}%
\end{equation}
as is indeed approriate for nonrelativistic physics. \ This result in Eq. (39)
combined with the basic expression for the rate of change of mechanical energy
in Eq. (36) then forces us to choose the nonrelativistic expression for
mechanical energy%
\begin{equation}
U_{m}=\frac{1}{2}m\left(  \frac{dx}{dt}\right)  ^{2}\text{ \ \ \ \ }%
U_{M}=\frac{1}{2}M\left(  \frac{dX}{dt}\right)  ^{2}%
\end{equation}
With these familiar nonrelativistic identifications, we find that the
conservation laws for linear momentum, angular momentum, energy, and constant
motion of the center of mass are all satisfied and the system is Galilean
invariant. \ There is no restriction on the force between the disks which is
given in Eq. (25).

Suppose now that we were to demand that our system of accelerating disks was
invariant under Lorentz transformation. \ This requires the result of Eq. (9)
which, from Eq. (33), becomes here%
\begin{align}
\frac{d}{dt}\left(  U\overrightarrow{\mathcal{X}}_{energy}\right)   &
=\widehat{i}\frac{d}{dt}\left(  U_{m}x+U_{M}X+\frac{C}{2-n}(X-x)^{2-n}%
\frac{(x+X)}{2}\right) \nonumber\\
&  =\widehat{i}\left[  \frac{dp_{m}}{dt}\frac{dx}{dt}x+U_{m}\frac{dx}%
{dt}+\frac{dp_{M}}{dt}\frac{dX}{dt}X+U_{m}\frac{dX}{dt}\right] \nonumber\\
&  +\widehat{i}\left[  C(X-x)^{1-n}\left(  \frac{dX}{dt}-\frac{dx}{dt}\right)
\frac{(x+X)}{2}+\frac{C}{2-n}(X-x)^{2-n}\frac{1}{2}\left(  \frac{dx}{dt}%
+\frac{dX}{dt}\right)  \right] \nonumber\\
&  =\widehat{i}\left(  \frac{dp_{m}}{dt}-C(X-x)^{1-n}\right)  \frac{dx}%
{dt}x+\widehat{i}\left(  \frac{dp_{M}}{dt}+C(X-x)^{1-n}\right)  \frac{dX}%
{dt}X+\widehat{i}\left[  U_{m}\frac{dx}{dt}+U_{m}\frac{dX}{dt}\right]
\nonumber\\
&  +\widehat{i}\left[  \frac{C(X-x)^{1-n}}{2}\left(  1-\frac{1}{2-n}\right)
\left(  \frac{dx}{dt}x-\frac{dX}{dt}X+\frac{dX}{dt}x-\frac{dx}{dt}X\right)
\right] \nonumber\\
&  =\widehat{i}\left[  U_{m}\frac{dx}{dt}+U_{m}\frac{dX}{dt}\right]
\nonumber\\
&  +\widehat{i}\left[  \frac{C(X-x)^{1-n}}{2}\left(  1-\frac{1}{2-n}\right)
\left(  \frac{dx}{dt}x-\frac{dX}{dt}X+\frac{dX}{dt}x-\frac{dx}{dt}X\right)
\right] \nonumber\\
&  =c^{2}\mathbf{P}%
\end{align}
where we have used the equations of motion in Eq. (28) to simplify the
expression. \ Thus Lorentz invariance for our system requires%
\begin{align}
c^{2}\mathbf{P}  &  \mathbf{=c}^{2}\mathbf{(}\widehat{i}p_{m}+\widehat{i}%
p_{M})\nonumber\\
&  =\widehat{i}U_{m}\frac{dx}{dt}+\widehat{i}U_{M}\frac{dX}{dt}\nonumber\\
&  +\widehat{i}\left[  \frac{C(X-x)^{1-n}}{2}\left(  1-\frac{1}{2-n}\right)
\left(  \frac{dx}{dt}x-\frac{dX}{dt}X+\frac{dX}{dt}x-\frac{dx}{dt}X\right)
\right]
\end{align}
Since the velocities $dx/dt,$ $dX/dt,$ and the positions $x$, $X$, are
arbitrary, the only way for this requirement to be met is for the mechanical
momentum $c^{2}\mathbf{p}_{mech}$ to be given by $U_{mech}d\mathbf{r/}dt$%
\begin{equation}
c^{2}\mathbf{p}_{mech}=U_{mech}d\mathbf{r/}dt
\end{equation}
and for the second line to vanish, implying%
\begin{equation}
\left(  1-\frac{1}{2-n}\right)  =0\text{ \ \ \ or \ \ \ \ }n=1
\end{equation}

Combining Eqs. (36) and (43), so as to eliminate the velocity $\mathbf{v=}%
d\mathbf{r}/dt$, we find $U_{mech}(dU_{mech}/dt)=c^{2}\mathbf{p}%
_{mech}\mathbf{\cdot(}d\mathbf{p}_{mech}/dt)$ so that $U_{mech}^{2}%
=c^{2}p_{mech}^{2}+const.$ \ Denoting this constant of integration by
$const=m^{2}c^{4},$ we have precisely the requirements of relativistic
mechanical momentum and energy related as%
\begin{equation}
U_{mech}=\left(  c^{2}p_{mech}^{2}+m^{2}c^{4}\right)  ^{1/2}%
\end{equation}
and comining this with Eq. (43), we find the familiar relativistic mechanical
momentum%
\begin{equation}
\mathbf{p}_{mech}=\frac{m\mathbf{v}}{(1-v^{2}/c^{2})^{1/2}}%
\end{equation}

The requirement in Eq. (44) that the exponent $n=1$ corresponds exactly to the
Coulomb potential in Eq. (23). \ Thus our disks provide a Lorentz-invariant
system only in the case where they can be reinterpreted within classical
electrodynamics as accelerating plates of a parallel-plate capacitor. \ We
should note that classical electrodynamics does indeed involve
velocity-dependent and acceleration dependent forces such as are mentioned in
Section D, but these forces do not enter the disk example within the
approximations $L<<R<<c^{2}/a$ where $a$ is the maximum acceleration of the plates.

The analysis here suggests three important aspects. \ First the conservation
laws of energy, linear momentum, and angular momentum can hold independent of
whether relativistic or nonrelativistic physics (or some combination of both)
is employed in the analysis. \ Second, Galilean invariance requires that
nonrelativistic expressions are used for mechanical energy and momentum but
makes no restrictions upon a potential function $V(|\mathbf{r}_{1}%
-\mathbf{r}_{2}|).$ \ Third, relativistic invariance requires not only that
relativistic expressions are used for the mechanical energy and momentum but
also places restrictions on the form of the interactions between the
particles. \ In the disk example, the Coulomb potential is the unique
potential associated with Lorentz invariance.

\subsection{Requirements for a Fully Relativistic Extension}

Although the calculations in Section C show that a system of two point
particles interacting through a potential function $V(|\mathbf{r}%
_{1}-\mathbf{r}_{2}|)$ can indeed be extended to a Lagrangian system which is
Lorentz invariant through order $v^{2}/c^{2},$ this, in general, is as far as
we can go. \ Already in equations (15) and (16) above, we have seen that a
relativistic system requires that the interaction between particles involves
not only energy $V(|\mathbf{r}_{1}-\mathbf{r}_{2}|)$ but also additional
velocity-dependent terms associated with the potential energy. \ Also, the
Lagrangian equations of motion involve an additional time derivative and so
require that there are velocity- and acceleration-dependent forces, as seen in
Eq. (19). \ Presumably a fully Lorentz-invariant interaction requires a full
field theory, and not every mechanical potential can be extended to a field
theory. \ 

For the Lorentz-invariant interaction of point particles, we expect the forces
to be transmitted at the speed of light $c.$ \ This speed is the only one
which is the same in every inertial frame. \ Thus we expect the forces to be
associated with a wave equation involving wave speed $c.$ \ This wave-equation
assumption has strong implications.

Let us consider the situation where the potential function arises from the
interaction of a very massive point particle at the origin of the $S$ frame
and a much lighter point particle $m$ at position $\mathbf{r}$, so that the
potential function can be regarded as given by $V(r)$ where $r=|\mathbf{r|}%
=(x^{2}+y^{2}+z^{2})^{1/2}$ is the distance from the origin in $S$. \ This
same situation can be observed from the $S^{\prime}$ frame moving with
constant velocity $\mathbf{u=}\widehat{\mathbf{i}}u$ relative to the $S$
frame. \ Then if $V(r)$ satisfies tensor behavior, we expect that in
$S^{\prime}$ the potential function $V^{\prime}(x^{\prime},y^{\prime
},z^{\prime},t^{\prime})$ moves rigidly with velocity $-\mathbf{u}%
=-\widehat{\mathbf{i}}u$ becoming a function of $x^{\prime}+ut^{\prime},$ and
so satisfies the wave equation%
\begin{equation}
\frac{1}{u^{2}}\frac{\partial^{2}V^{\prime}}{\partial t^{\prime2}}%
-\frac{\partial^{2}V^{\prime}}{\partial x^{\prime2}}=(\delta-\text{function
singularity at }\mathbf{r}^{\prime}=-\mathbf{u}t^{\prime})
\end{equation}
Furthermore, the relativistic behavior requires that the potential function
$V^{\prime}(x^{\prime},y^{\prime},z^{\prime},t^{\prime})$ acting on the
particle $m$ arises from a signal traveling with velocity $c,$ and so the
potential function satisfies the wave equation%
\begin{equation}
\frac{1}{c^{2}}\frac{\partial^{2}V^{\prime}}{\partial t^{\prime2}}%
-\frac{\partial^{2}V^{\prime}}{\partial x^{\prime2}}-\frac{\partial
^{2}V^{\prime}}{\partial y^{\prime2}}-\frac{\partial^{2}V^{\prime}}{\partial
z^{\prime2}}=(\delta-\text{function singularity at }\mathbf{r}^{\prime
}=-\mathbf{u}t^{\prime})
\end{equation}
Subtracting Eq. (47) from Eq. (48) so as to eliminate the time derivatives, we
find%
\begin{equation}
c^{2}\left(  1-\frac{u^{2}}{c^{2}}\right)  \frac{\partial^{2}V^{\prime}%
}{\partial x^{\prime2}}+c^{2}\frac{\partial^{2}V^{\prime}}{\partial
y^{\prime2}}+c^{2}\frac{\partial^{2}V^{\prime}}{\partial z^{\prime2}}%
=(\delta-\text{function singularity at }\mathbf{r}^{\prime}=-\mathbf{u}%
t^{\prime})
\end{equation}
This suggests Lorentz contraction in the $x^{\prime}$-direction. \ Also, if we
take the limit as $u$ goes to zero so that we are back in the $S$ frame where
the potential function is time independent, then we find Eq. (49) becomes%
\begin{equation}
\frac{\partial^{2}V}{\partial x^{2}}+\frac{\partial^{2}V}{\partial y^{2}%
}+\frac{\partial^{2}V}{\partial z^{2}}=(\delta-\text{function singularity at
}r=0)
\end{equation}
Thus the potential $V(r)$ which allows both the natural rigid behavior in
another inertial frame and a natural relativistic extension to wave behavior
at the relativistic speed $c$ must necessarily satisfy Laplace's equation.
\ But if the potential is rotationally symmetric in the $S$ frame where the
massive particle is at rest, then the potential satisfying Laplace's equation
must be the Coulomb potential%
\begin{equation}
V(r)=\frac{k}{r}%
\end{equation}
We conclude that a nonrelativistic potential function arising from a point
source which allows a natural extension to a relativistic theory involving the
wave equation must necessarily be the Coulomb potential.

The calculations given here suggest the possiblity that the nonrelativistic
Lagrangian for the interaction of two point particles given in Eq. (8) may be
an approximation to nature only in the case of the Coulomb/Kepler
potential.\cite{Yukawa} \ In the case of interacting electric charges, the
extension to a relativistic interaction through order $v^{2}/c^{2}$ gives the
Darwin Lagrangian. \ And the Darwin Lagrangian is known to be a valid
approximation to fully relativistic classical electrodynamics.

\subsection{Implications for Classical Physics}

Classical electromagnetism is a relativistic theory which was developed during
the nineteenth century before the ideas of special relativity. \ Indeed,
special relativity arose at the beginning of the twentieth century as a
response to the conflict of electromagnetism with nonrelativistic mechanics.
\ Around the same time, quantum mechanics was introduced in response to the
mismatch between electromagnetic radiation equilibrium (blackbody radiation)
and classical statistical mechanics (which is based on nonrelativistic
mechanics). \ Although quantum theory and special relativity have gone on to
enormous successes, they have left behind a number of unresolved questions
within classical physics. \ For example, the blackbody radiation problem has
never been solved within relativistic classical physics.\cite{bl} There have
been discussions of classical radiation equilibrium using nonrelativistic
mechanical scatterers and even one calculation of a scattering particle using
relativistic mechanical momentum in a general class of non-Coulomb
potentials.\cite{Blanco} \ However, there has never been a treatment of
scattering by a relativistic particle in a Coulomb potential, despite the fact
that the Coulomb potential has all the qualitative aspects which might allow
classical radiation equilibrium at a spectrum with finite thermal energy.

We conclude that the misconceptions regarding potentials which allow
extensions to relativistic systems is relevant for treatments in mechanics
textbooks and perhaps also for the description of nature within classical theory.

\bigskip

Acknowledgement

The argument given here in Section E was adapted from the work of Dr. Hans de
Vries appearing as the answer to an unrelated query on the internet, http://www.physicsforums.com/showthread.php?t=114620.


\begin{thebibliography}{99}                                                                                               %


\bibitem {Goldstein}H. Goldstein, C. Poole, and J. Safko, \textit{Classical
Mechanics 3rd ed. }(Addison-Wesley, New York 2002), p. 313.

\bibitem {Jose}J. V. Jose and E.J. Saletan, \textit{Classical Dynamics: A
Contemporary Approach} (Cambridge University Press 1998), p. 210. \ The text
includes some remarks indicating discomfort with the "relativistic Lagrangian."

\bibitem {Goldstein2}See ref. 1, Section 7.9, p. 316.

\bibitem {Brehme}One referee for an earlier version of this article wrote, "My
understanding of what is meant by the 'relativistic Lagrangian' is a scalar
functional of spacetime coordinates and their proper time derivatives,
$L=L(x^{\mu},dx^{\mu}/d\tau),$ that is invariant under all coordinate
transformations in a Minkowski spacetime. This is the definition used by A. O.
Barut in his monograph, \textit{Electrodynamics and Classical Theory of Fields
and Particles}, MacMillan, 1964." \ This is the definition used by R.W.
Brehme, "The Relativistic Lagrangian," Am. J. Phys. \textbf{39, }275-280 (1971).

\bibitem {Blanco}This mistaken idea of Lorentz-invariant behavior as a
description of nature involving only the use of relativistic mechanical
momentum appears in the classical theoretical analysis of blackbody radiation
by R. Blanco, L. Pesquera, and E. Santos, "Equilibrium between radiation and
matter for classical relativistic multiperiodic systems. Derivation of
Maxwell-Boltzmann distribution from Rayleigh-Jeans spectrum," Phys. Rev. D
\textbf{27}, 1254-1287 (1983), and "Equilibrium between radiation and matter
for classical relativistic multiperiodic systems. II. Study of radiative
equilibrium with Rayleigh-Jeans radiaiton," \textit{ibid}. \textbf{29},
2240-2254 (1984).

\bibitem {Kretchman}E. Kretchmann, "\"{U}ber den physikalischen Sinn der
Relativit\"{a}tspostulate: A. Einsteins neue und seine urspr\"{u}ngliche
Relativit\"{a}tstheorie," Annalen der Physik \textbf{53}, 575-614 (1917).

\bibitem {CV}S. Coleman and J.H. Van Vleck, "Origin of 'hidden momentum
forces' on magnets," Phys. Rev. \textbf{171}, 1370-1375 (1968).

\bibitem {ref12}See, for example, ref. 1, p. 313, Eq. (7.136) or ref. 2, p.
210, Eq. (5.27).

\bibitem {CofE}See, for example, ref. 7 or T. H. Boyer, "Illustrations of the
relativistic conservation law for the center of energy," Am. J. Phys.
\textbf{73}, 953-961 (2005).

\bibitem {Jack}I am not aware of a derivation of the Darwin Lagrangian which
takes the form given here. \ A different derivation is given by J. D. Jackson,
\textit{Classical Electrodynamics 3rd ed} (Wiley, New York 1999), p. 596-598;
the Darwin Lagrangian is given in Eq. (12.82).

\bibitem {PA}The expressions for the electric and magnetic fields agree with
those given by L. Page and N.I. Adams, "Action and Reaction Between Moving
Charges," Am. J. Phys. \textbf{13}, 141-147\ (1945).

\bibitem {Illustrations}See the article by Boyer in reference 9.

\bibitem {Coul12}See, for example, ref. 2, pp. 211-212.

\bibitem {Jack3}See, for example, ref. 10, p. 582, Eq. (12.12).

\bibitem {Yukawa}Some people have suggested that particle interactions through
the Yukawa potential provide a relativistic interaction; however, I am not
aware of any \textit{classical} relativistic theory of such forces.

\bibitem {bl}See, for example, the discussions by T. H. Boyer, "Blackbody
radiation, conformal symmetry, and the mismatch between classical mechanics
and electromagnetism," J. Phys. A: Math. Gen. \textbf{38}, 1807-1821 (2005);
"Connecting blackbody radiation, relativity, and discrete charge in classical
electrodynamics," Found. Phys. \textbf{37}, 999-1026 (2007).
\end{thebibliography}
\end{document}